\documentclass[aps,prb,twocolumn,showpacs,amsmath,amssymb,superscriptaddress]{revtex4}
\pdfoutput=1
\usepackage{graphicx}
\usepackage{amssymb}
\usepackage{natbib}
\usepackage{calc}
\usepackage{SIunits}
\usepackage{textcomp}
\usepackage{color}

\begin{document}

\title{Spin Excitations and Spin Wave Gap in the Ferromagnetic Weyl Semimetal Co$_3$Sn$_2$S$_2$}

\author{Chang Liu}
\thanks{These authors made equal contributions to this paper}
\affiliation{Beijing National Laboratory for Condensed Matter
Physics, Institute of Physics, Chinese Academy of Sciences, Beijing
100190, China}
\affiliation{School of Physical Sciences, University of Chinese Academy of Sciences, Beijing 100190, China}
\author{Jianlei Shen}
\thanks{These authors made equal contributions to this paper}
\affiliation{Beijing National Laboratory for Condensed Matter
Physics, Institute of Physics, Chinese Academy of Sciences, Beijing
100190, China}
\affiliation{School of Physical Sciences, University of Chinese Academy of Sciences, Beijing 100190, China}
\author{Jiacheng Gao}
\thanks{These authors made equal contributions to this paper}
\affiliation{Beijing National Laboratory for Condensed Matter
Physics, Institute of Physics, Chinese Academy of Sciences, Beijing
100190, China}
\affiliation{School of Physical Sciences, University of Chinese Academy of Sciences, Beijing 100190, China}
\author{Changjiang Yi}
\affiliation{Beijing National Laboratory for Condensed Matter
Physics, Institute of Physics, Chinese Academy of Sciences, Beijing
100190, China}
\affiliation{School of Physical Sciences, University of Chinese Academy of Sciences, Beijing 100190, China}
\author{Di Liu}
\affiliation{Beijing National Laboratory for Condensed Matter
Physics, Institute of Physics, Chinese Academy of Sciences, Beijing
100190, China}
\affiliation{School of Physical Sciences, University of Chinese Academy of Sciences, Beijing 100190, China}
\author{Tao Xie}
\affiliation{Beijing National Laboratory for Condensed Matter
Physics, Institute of Physics, Chinese Academy of Sciences, Beijing
100190, China}
\affiliation{School of Physical Sciences, University of Chinese Academy of Sciences, Beijing 100190, China}
\author{Lin Yang}
\affiliation{Beijing National Laboratory for Condensed Matter
Physics, Institute of Physics, Chinese Academy of Sciences, Beijing
100190, China}
\affiliation{School of Physical Sciences, University of Chinese Academy of Sciences, Beijing 100190, China}
\author{Sergey Danilkin}
\affiliation{Australian Centre for Neutron Scattering, Australian Nuclear Science and
Technology Organization, Lucas Heights NSW-2234, Australia}
\author{Guochu Deng}
\affiliation{Australian Centre for Neutron Scattering, Australian Nuclear Science and
Technology Organization, Lucas Heights NSW-2234, Australia}
\author{Wenhong Wang}
\affiliation{Beijing National Laboratory for Condensed Matter
Physics, Institute of Physics, Chinese Academy of Sciences, Beijing
100190, China}
\affiliation{Songshan Lake Materials Laboratory, Dongguan, Guangdong 523808, China }
\author{Shiliang Li}
\affiliation{Beijing National Laboratory for Condensed Matter
Physics, Institute of Physics, Chinese Academy of Sciences, Beijing
100190, China}
\affiliation{School of Physical Sciences, University of Chinese Academy of Sciences, Beijing 100190, China}
\affiliation{Songshan Lake Materials Laboratory, Dongguan, Guangdong 523808, China }
\author{Youguo Shi}
\affiliation{Beijing National Laboratory for Condensed Matter
Physics, Institute of Physics, Chinese Academy of Sciences, Beijing
100190, China}
\affiliation{School of Physical Sciences, University of Chinese Academy of Sciences, Beijing 100190, China}
\affiliation{Songshan Lake Materials Laboratory, Dongguan, Guangdong 523808, China }
\affiliation{Physical Science Laboratory, Huairou National Comprehensive Science Center, Beijing, China}
\author{Hongming Weng}
\email{hmweng@iphy.ac.cn}
\affiliation{Beijing National Laboratory for Condensed Matter
Physics, Institute of Physics, Chinese Academy of Sciences, Beijing
100190, China}
\affiliation{School of Physical Sciences, University of Chinese Academy of Sciences, Beijing 100190, China}
\affiliation{Songshan Lake Materials Laboratory, Dongguan, Guangdong 523808, China }
\affiliation{Physical Science Laboratory, Huairou National Comprehensive Science Center, Beijing, China}
\author{Enke Liu}
\email{ekliu@iphy.ac.cn}
\affiliation{Beijing National Laboratory for Condensed Matter
Physics, Institute of Physics, Chinese Academy of Sciences, Beijing
100190, China}
\affiliation{Songshan Lake Materials Laboratory, Dongguan, Guangdong 523808, China }
\author{Huiqian Luo}
\email{hqluo@iphy.ac.cn}
\affiliation{Beijing National Laboratory for Condensed Matter
Physics, Institute of Physics, Chinese Academy of Sciences, Beijing
100190, China}
\affiliation{Songshan Lake Materials Laboratory, Dongguan, Guangdong 523808, China }

\date{\today}

\begin{abstract}
We report a comprehensive neutron scattering study on the spin excitations in the magnetic Weyl semimetal Co$_3$Sn$_2$S$_2$ with quasi-two-dimensional structure.  Both in-plane and out-of-plane dispersions of the spin waves are revealed in the ferromagnetic state, similarly dispersive but damped spin excitations persist into the paramagnetic state. The effective exchange interactions have been estimated by a semi-classical Heisenberg model to consistently reproduce the experimental $T_C$ and spin stiffness. However, a full spin wave gap below $E_g=2.3$ meV is observed at $T=4$ K, much larger than the estimated magnetic anisotropy energy ($\sim0.6$ meV), while its temperature dependence indicates a significant contribution from the Weyl fermions. These results suggest that Co$_3$Sn$_2$S$_2$ is a three-dimensional correlated system with large spin stiffness, and the low-energy spin dynamics could interplay with the topological electron states.

\end{abstract}

\maketitle
\section{Introduction}

Magnetic topological materials, which combine non-trivial band topology and magnetic order, have great potential on fundamental physics and technology applications due to a number of exotic quantum phenomena such as quantum anomalous Hall effect, topological axion state and chiral Majorana fermions, etc \cite{bhyan2017,ytokura2017,gxu2019,hmweng2015,cshekhar2015,qlhe2017}. In recent years, more and more intrinsic magnetic materials have been theoretically predicted as magnetic Dirac semimetals (DSMs), Weyl semimetals (WSMs) and topological insulators (TIs) \cite{hchen2014,ptang2016,zjwang2016,gqchang2016,qnxu2018,gxu2018,ymshi2019,dqzhang2019}. Though a few of them were experimentally confirmed for now \cite{jyliu2017,lye2018,ekliu2018,qwang2018,dfliu2019,nmorali2019,hmweng2019,ibelopolski2019,motrokov2019,ybzhang2020}, their spin dynamics and the interplay with topological electron states are far away from being well understood \cite{ppark2018,sitoh2013}.

Specifically in ferromagnetic WSMs, the Weyl nodes serve as the magnetic monopoles of the Berry curvature, which leads to the intrinsic anomalous Hall effect (AHE) in bulk transport properties \cite{fhaldane2004,dxiao2010,aaburkov2014}. At the same time, these Weyl nodes can also affect the spin wave dispersions, since around Weyl nodes the spin operators and the current operators are in one-to-one correspondence, which builds up the direct connection between spin dynamics and AHE \cite{sitoh2016,monoda2008}. For example, in a magnetic WSM candidate SrRuO$_3$, the impact of Weyl points induces a nonmonotonous temperature dependence of the anomalous Hall conductivity (AHC) $\sigma_{xy}(T)$, which can be also traced both in the spin wave gap $E_g$ and its stiffness $D$ \cite{sitoh2016,kjenni2019}. Such results are distinct from the conventional ferromagnetic metals, where their spin waves are usually either gapless in the weak correlation limit, or show a monotonous spin gap following the magnetic order parameter in the strong correlation limit with spin-orbit coupling (SOC)  \cite{monoda2008}. Careful investigations on the energy dependence of spin excitations below and above the Curie temperature ($T_C$) can provide important information about the spin-spin interactions and correlations concerning their crucial roles in the AHE.

Co$_3$Sn$_2$S$_2$ is a new experimentally verified ferromagnetic WSM with very promising topological properties \cite{ekliu2018,qwang2018,dfliu2019,nmorali2019,hmweng2019}.  It is a Shandite compound in a rhombohedral structure (space group: $R\overline{3}m$) with quasi-two-dimensional (quasi-2D) Co$_3$Sn layers sandwiched between S atoms  \cite{rweihrich2005,rweihrich2006}. The magnetic Co atoms arrange on a kagome lattice in the ab-plane with about 0.3 $\mu_B$/Co ordered moments aligned along $c-$axis below $T_C=$ 177 K [Fig. 1(a)] \cite{pvaqueiro2009,wschnelle2013,makassem2016b}. There are three pairs of Weyl nodes in the first Brillouin zone close to Fermi level [Fig. 1(b)], as evidenced by the surface Fermi-arcs and linear bulk band dispersions from spectroscopic experiments \cite{ekliu2018,qwang2018,qnxu2018,dfliu2019,nmorali2019}. With the considerably enhanced Berry curvature from its band structure [Fig. 1(c)], a record of large AHC is found up to 1,130 $\Omega^{-1}$cm$^{-1}$ with strong temperature dependence approaching $T_C$ \cite{ekliu2018,ryang2020}.

Here in this paper, we report a comprehensive neutron scattering study on the Co$_3$Sn$_2$S$_2$ single crystals by measuring the spin excitations up to 18 meV both in the ferromagnetic and paramagnetic states.  Both in-plane and out-of-plane dispersions are found, suggesting the magnetic interactions are actually three-dimensional (3D) in despite of its quasi-2D lattice structure. The paramagnetic excitations above $T_C$ ($T=200$ K) show similar dispersions but damped intensities. Theoretical calculations on the effective exchange couplings well reproduce the experimental $T_C$ and large spin stiffness, but give a magnetic anisotropy energy ($\sim0.6$ meV) much smaller than the spin wave gap $E_g=$2.3 meV observed at $T=4$ K. Further analysis on the temperature dependence of the gap suggests a significant contribution from AHC. Therefore, Co$_3$Sn$_2$S$_2$ is a moderately correlated ferromagnet, where the conducting electrons related to Weyl fermions are deeply involved into its spin dynamics.

\begin{figure}[t]
\includegraphics[width=0.45\textwidth]{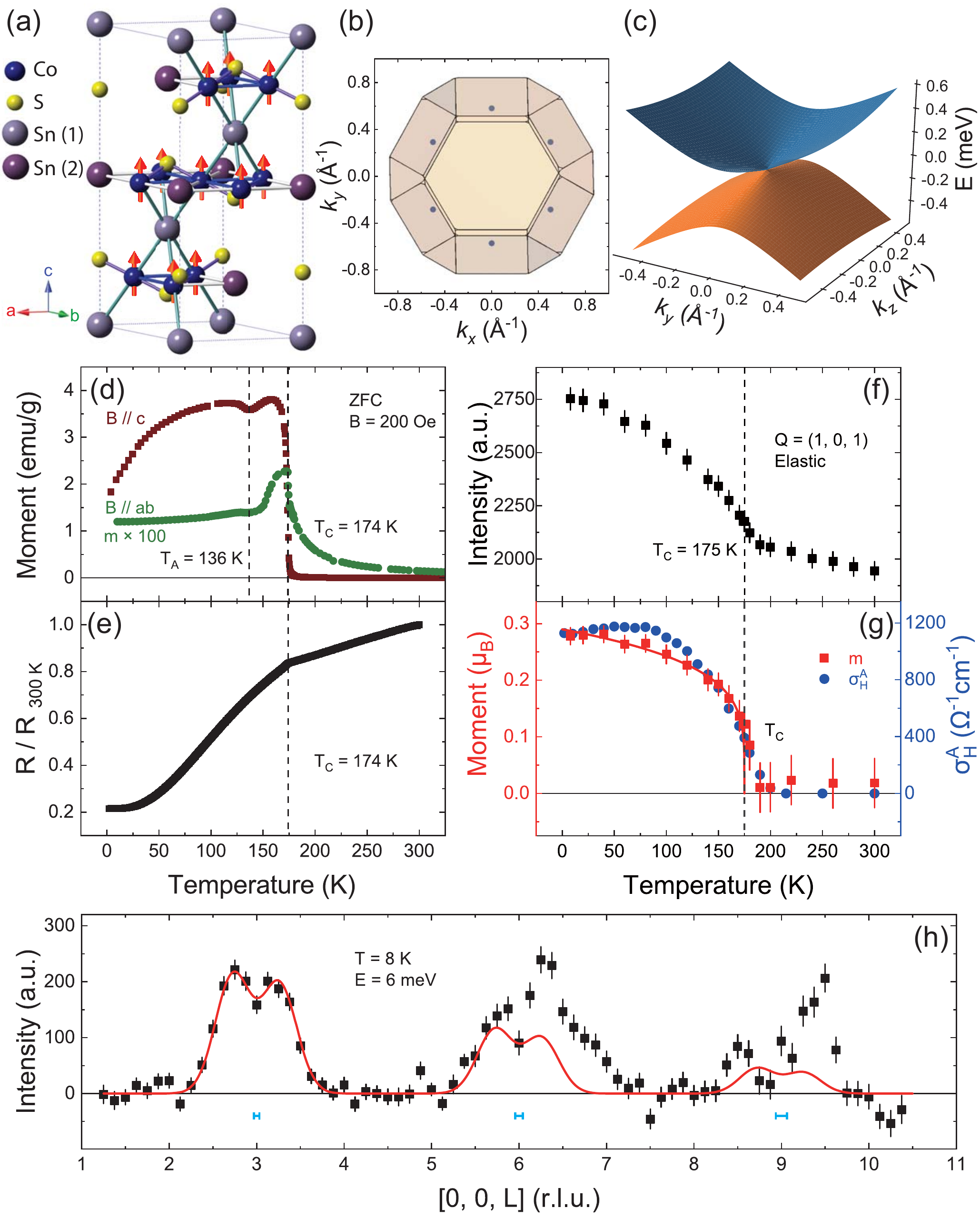}
\caption{
(a) Crystal and magnetic structure in Co$_3$Sn$_2$S$_2$. (b) The location of Weyl points in Brillouin zone along the $z$ axis. (c) The
dispersion of Weyl Hamiltonian on the $k_x=0$ plane. (d) and (e) Temperature dependence of magnetization and resistivity. (f) and (g) Temperature dependence of the peak intensity at $Q= (1, 0, 1)$, and the deduced ordered moment $m$ in comparison with the anomalous Hall conductivity $\sigma^A_H$. (h) Spin excitations at 6 meV along [0, 0, L] direction. The red lines are two-gaussian-fitting based on $L=3$ data and normalized by the magnetic form factor for $L=6$ and 9. The horizontal bars mark the calculated instrument resolution.
}
\end{figure}

\section{Experiment details}

We prepared high quality Co$_3$Sn$_2$S$_2$ single crystals using the flux method as reported before (See Supplemental Materials). The zero-field-cooling magnetization and resistivity show a clear ferromagnetic transition at $T_C=$ 174 K, and another anomaly at $T_A=$ 136 K [Fig. 1(d)(e)], which may be related to the forming of another in-plane antiferromagnetic order \cite{makassem2017,zguguchia2019}. The huge anisotropy of magnetization between $B \parallel c$ and $B \parallel ab$ geometry confirms the $c-$axis polarized magnetism \cite{pvaqueiro2009,wschnelle2013,makassem2016b}. Neutron scattering experiments were carried out using a thermal triple-axis spectrometer Taipan and a cold triple-axis spectrometer Sika at Australian Centre for Neutron Scattering, ANSTO, Australia \cite{sdanilkin2009,cmwu2016}. For Taipan experiment, the final neutron energy was fixed as $E_f=$ 14.87 meV, with a pyrolytic graphite filter, a double focusing  monochromator and a vertical focusing analyzer. About 1.5 grams single crystals of Co$_3$Sn$_2$S$_2$ were co-aligned by hydrogen-free glue on several thin aluminum plates. For Sika experiment, the final energy was chosen as $E_f=$ 5 meV with a cooled Be filter, a double focusing  monochromator and a flat analyzer. A large piece of single crystal with mass about 7.6 grams was used. We defined the scattering plane $[H, 0, 0] \times [0, 0, L]$ using hexagonal unit cell where $a=b= 5.352$ \AA, $c=13.095$ \AA, $\alpha=\beta=90^{\circ}$, $\gamma=120^{\circ}$, and $\textbf{Q}=H\textbf{a} ^*+K\textbf{b} ^*+L\textbf{c} ^*$, where $H$, $K$, and $L$ are Miller indices. Thus the $d-$spacing is given by: $d_{HKL}=1/\sqrt{4(H^2+HK+K^2)/3a^2+L^2/c^2}$ \cite{gcwang2014}. The instrument resolution is calculated by ResLib \cite{reslib}.

\section{Results and discussion}

\subsection{Magnetic order parameter}

We have first performed elastic neutron scattering measurements on a Bragg peak at $Q=$ (1, 0, 1), where the nuclear scattering is weak. The integrated intensity of (1, 0, 1) peak shows a clear ferromagnetic transition at $T_C=$ 175 K, consistent with the magnetization results [Fig. 1(f)]. The ordered moment $M$ can be estimated by comparing the intensity between the magnetic scattering and nuclear scattering \cite{dlgong2018}(See Supplemental Materials), giving $0.28\pm0.02$ $\mu_B$ per Co atom at the base temperature $T=8$ K. The weak ordered moment is consistent with previous report on powder neutron diffraction experiments \cite{pvaqueiro2009} and reflects its itinerant and semi-metallic character, where both holes and electrons contribute to the Fermi surfaces \cite{ekliu2018,qwang2018,qnxu2018}. The magnetic order parameter $M(T)$ follows similar temperature dependence as AHC except a discrepancy from $T=50$ K to 150 K, because the intrinsic AHC is mostly determined by the Berry curvature at low temperatures away from $T_C$ [Fig. 1(g)] \cite{ekliu2018}. The critical exponent $\beta$ is determined to be $0.21\pm0.04$ by fitting the magnetic order parameter using $M(T)=M_0(1-T/T_C)^{\beta}$ for a second order magnetic transition [red line in Fig. 1(g)].

\begin{figure}[t]
\includegraphics[width=0.45\textwidth]{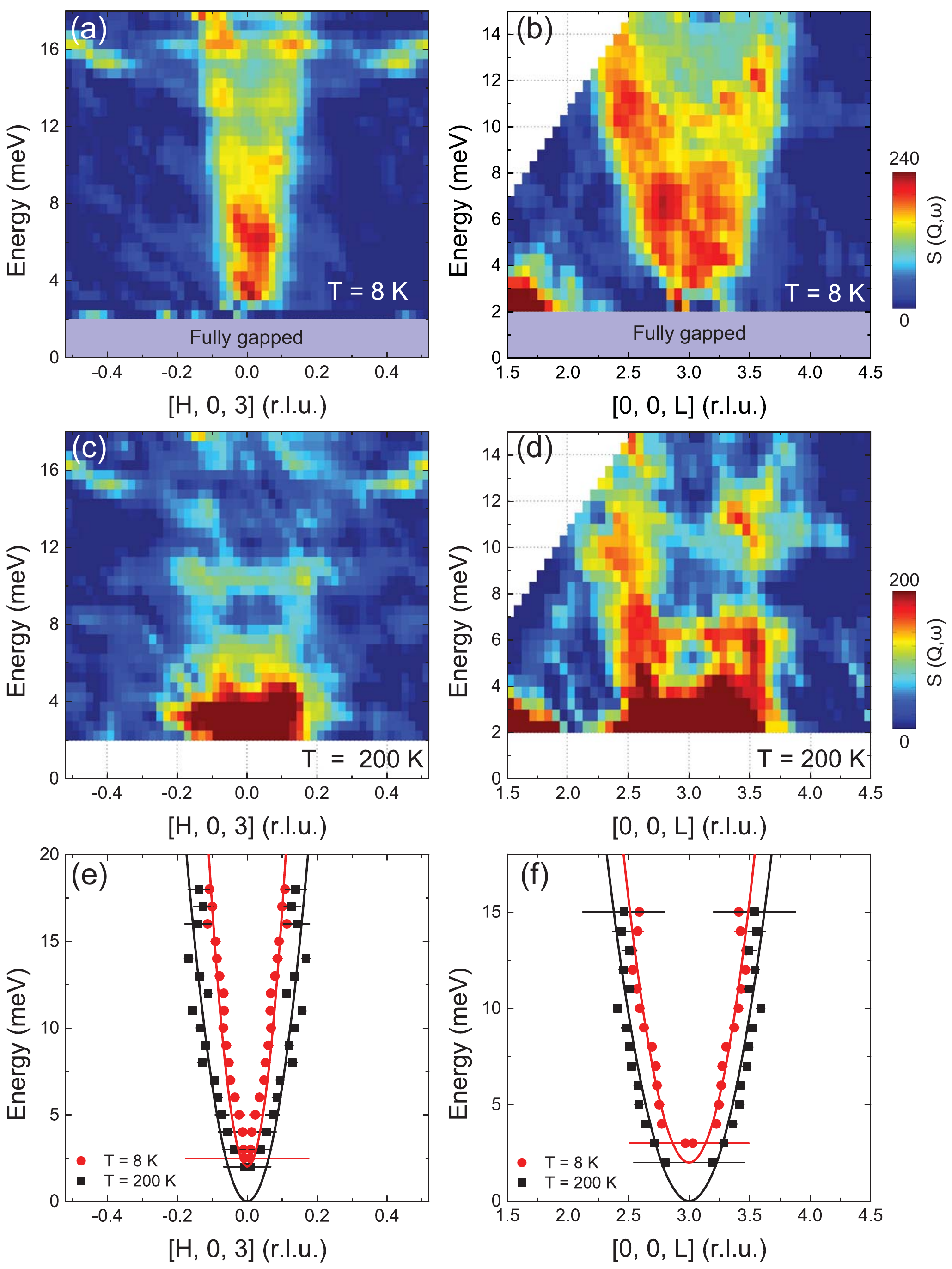}
\caption{(a)-(d) In-plane and out-of-plane spin waves at $T= 8$ K and paramagnetic excitations at $T= 200$ K around $Q=$ (0, 0, 3). The high energy data at low $Q$ side in panel (c) and (d) is missing due to the scattering restriction, and the data below 2 meV is not measured for strong contamination from quasi-elastic scattering.
(e) and (f) Dispersion of the spin excitations obtained from two-gaussian-fitting of the raw data. The solids lines are fitting results with $q^2-$dependence.
 }
 \end{figure}

\begin{figure}[t]
\includegraphics[width=0.45\textwidth]{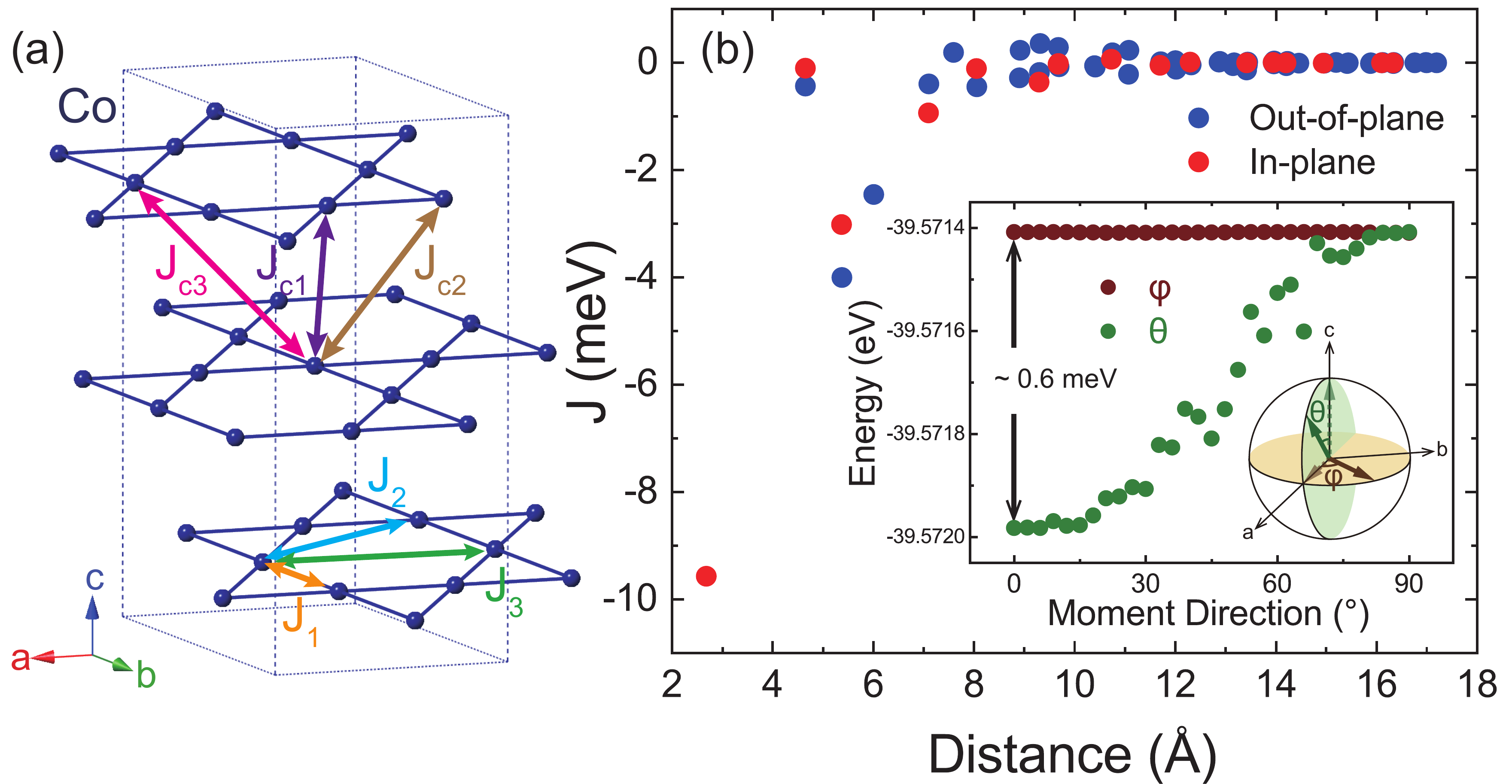}
\caption{(a) Effective exchange interactions between Co atoms in our first principle calculations. (b) Estimation of the effective correlation length of the exchange interaction energies (main panel) and the total energy for the moment rotation (insert) for both in-plane and out-of-plane cases.
 }
\end{figure}

\subsection{Low-energy spin waves}

 As neutron scattering cannot actually approach the primary ferromagnetic excitations centering around $Q=0$ at finite energy transfer \cite{jvkranendonk1958}, we instead seek the magnetic excitation signal along $Q=$ (0, 0, L), and measure the background at $Q=$ ($\pm0.3$, 0, L) or ($\pm0.5$, 0, L) at identical energies. Fig. 1(h) shows a typical constant-energy scan at $E=6$ meV and $T=8$ K. Apparently the signals only locally show up around $L=3, 6, 9$ with two splitting incommensurate peaks, where signals at high $Q$ are contaminated by the phonon scattering from Co$_3$Sn$_2$S$_2$ itself and the aluminium sample holder. We have further mapped out the spin waves by constant-energy scans from $E=2$ meV to 18 meV along both $[H, 0, 3]$ and $[0, 0, L]$ directions at $T=8$ K [Fig. 2(a) and (b)]. Since there are no peaks centering $Q=(0, 0, 3)$ both along $H$ and $L$ directions at 2 meV, we tend to consider a full spin wave gap below 2 meV even it cuts the resolution edge of a thermal triple-axis spectrometer. Above 2 meV, a steep dispersion of the in-plane spin waves along $H$ direction is well established, while the out-of-plane spin waves along $L$ direction are more dispersive and broad in peak width. The phonon signals also show up outside the spin wave branches around 16 meV (See Supplemental Materials). By warming up to the paramagnetic state at $T=200$ K, the spin excitations above 8 meV are heavily damped, and the low-energy spin excitations below 4 meV are stronger, as the gap is closed [Fig. 2(c) and (d)]. To qualitatively compare the dispersion below and above $T_C$, we have carried out two-gaussian-fitting on the raw data after subtracting the background accordingly (See Supplemental Materials). Because we are unable to reach the band top at zone boundary for scattering limitations, a full fitting based on a microscopic model of spin waves is impossible by these limited data. However, we can still roughly fit the dispersion at 8 K by the $q^2-$dependence: $E=E_g+Dq^2$ (leading to a spin wave gap at 2 meV), where $D\sim S\mid J \mid$ represents the spin stiffness in a ferromagnetic system with exchange coupling $J$ and effective spin $S$ [Fig. 2(e) and (f)]\cite{dprice2015}. The spin excitations at 200 K is gapless and can be also fitted by same equation with smaller $D$. We summarize the fitting results of $D$ in Table I, both in ferromagnetic and paramagnetic states $D$ is quite large in comparison with that in SrRuO$_3$ \cite{sitoh2016,kjenni2019}.

\begin{table}
\caption{Experimental and calculated results of $D$ and $T_C$}
\label{Tab.1}
\begin{centering}
{
\begin{tabular}{|c|c|c|c|c|}
\hline\hline
Experiment & $D_H(8 K)$     & $D_L (8 K)$    &$D_H(200 K)$  &$D_L(200 K)$ \\
 & (meV\AA$^2$) & (meV\AA$^2$) & (meV\AA$^2$) & (meV\AA$^2$)  \\
\hline
      &$803\pm46$      &$237\pm13$  &$360\pm30$  &$169\pm10$\\
\hline
Calculation & $D_{xx}$     & $D_{yy}$    &$D_{zz}$  & $T_C$\\
 & (meV\AA$^2$) & (meV\AA$^2$) & (meV\AA$^2$) & (K) \\
\hline
      &$945$      &$833$  &$656$  & 167 \\
\hline\hline
\end{tabular}
}
\end{centering}
\end{table}

\begin{figure}[t]
\includegraphics[width=0.45\textwidth]{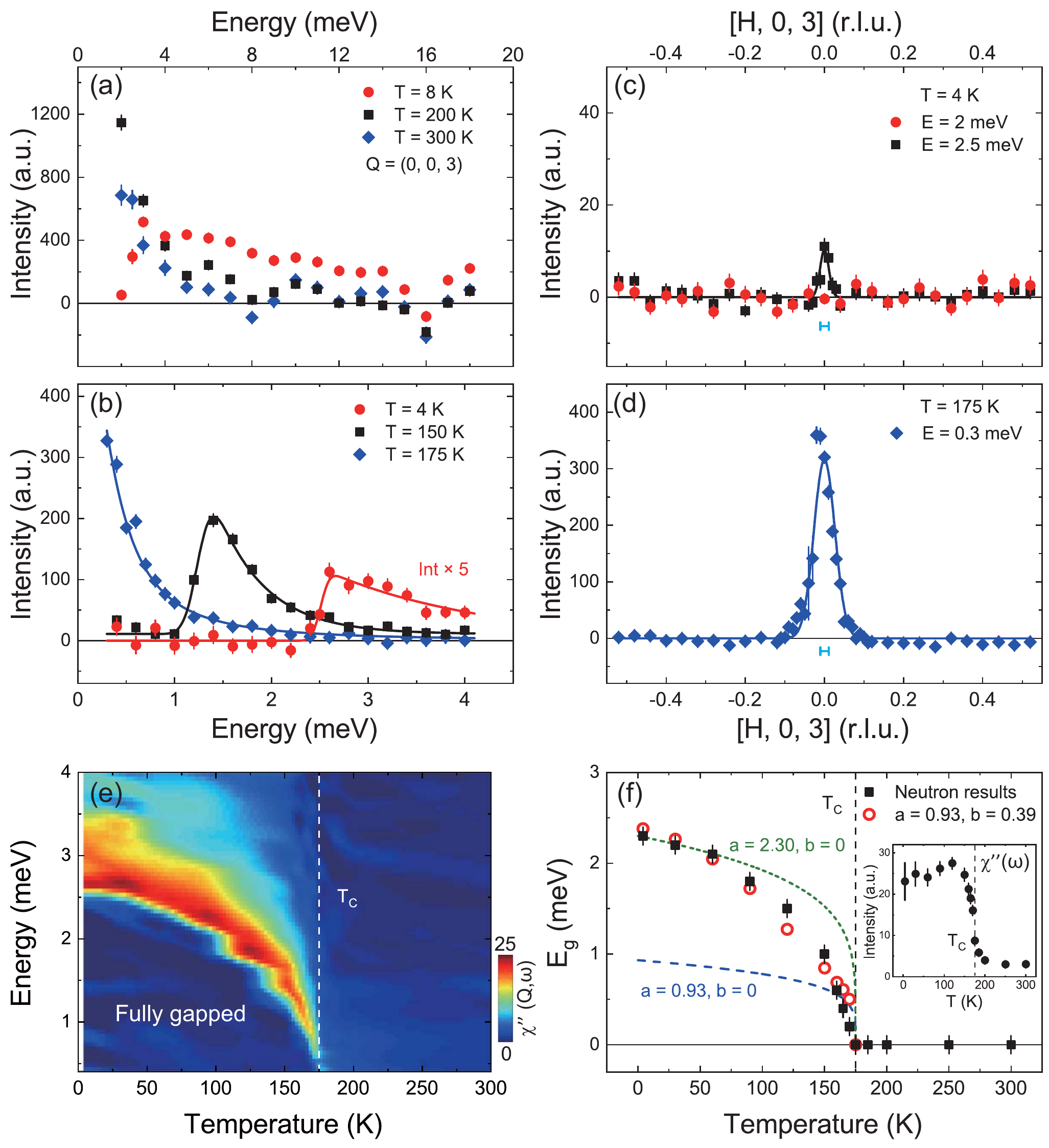}
\caption{(a) Energy dependence of the spin excitations at $T=$ 8, 200 and 300 K measured at Taipan. (b) High resolution energy dependence of the spin excitations at $T=$ 4, 150 and 175 K measured at Sika. (c) and (d) Constant-energy scans for $E=2$ and 2.5 meV at $T=4$ K, and $E=0.3$ meV at $T=175$ K. The solid lines are gaussian fittings, and the horizontal bars mark the instrument resolution. (e) Temperature dependence of the dynamic susceptibility $\chi^{\prime\prime}(Q,\omega)$ at low energy. (f) Temperature dependence of the spin wave gap and fitting results by Eq.2 with different parameters. Insert shows the integral intensity $\chi^{\prime\prime}(\omega)$  up to 4 meV.
 }
\end{figure}

\subsection{Theoretical calculations}

Although the Co$_3$Sn$_2$S$_2$ compound is believed as an itinerant ferromagnet below $T_C$ \cite{ekliu2018,qwang2018}, we can still estimate the effective exchange couplings by viewing the ordered moments located on Co sites and then calculating the total energy variation for small deviations of ground state magnetic configuration. A semi-classical analogous to the Heisenberg Hamiltonian can be written as
\begin{equation}
  H = \sum_{<i,j>} J_{ij} \mathbf{e}_i \mathbf{e}_j.
\end{equation}
Here $\mathbf{e}_{i,j}$ are unit vectors point to the direction of the spin $S_{i,j}$, and $J_{ij}$ stands for the exchange coupling between the $i$th and $j$th sites, as shown in Fig.3 (a). Thus we can define the total exchange interactions between a central Co atom and all its neighbours up to quite long range where the exchange interaction is zero: $J_0 = \sum_{i} J_{0i}$. By implementing the calculation of spin interactions between one and 400 neighbouring Co atoms by using a multiple-scattering Green function method \cite{dmmkorotin2015}, we have obtained that $J_1= -2.37$ meV, $J_2= -0.16$ meV, $J_3= -0.02$ meV, $J_{c1}= -0.30$ meV, $J_{c2}= -0.75$ meV and $J_{c3}= -0.11$ meV (See Supplemental Materials). We have to admit such calculation is highly model dependent, however, the calculated results can be verified by experimental data. First, the Curie temperature is determined by: $T_C = 2J_0 \langle M^2\rangle /3k_B\langle M\rangle^2 $, where $\langle M^2\rangle /\langle M\rangle^2 = 1$ for a classical ferromagnetic system \cite{liechtenstein1987}. Thus we calculate $T_C = 2J_0/3k_B=167$ K, which is fairly close to the experimental value (175 K). Second, the spin stiffness tensor is determined by the following expression:
\begin{equation}
  D_{\alpha\beta} = \frac{2}{M} \sum_j J_{0j}R_{j\alpha}R_{j\beta},
\end{equation}
where $R_{j\alpha}$ means the $\alpha$-direction component of lattice vector ${\bf R}_{j}$. By using the above $J_{ij}$, we have calculated the in-plane spin stiffness and it agrees with experimental data, but the out-of-plane spin stiffness may be overestimated in such a local moment picture (Table I). Third, the comparable exchange couplings for out-of-plane ($J_{ci}$) and in-plane ($J_{1,2}$) also support the 3D spin waves, since the locations of ordered moments in Co chains along $c-$axis do not exactly match but shift along $a-$axis [Fig.1(a)]. We have further summed over all contributions from the atoms having the same distance to the central Co atom, and found all spin interactions have an effective distance about 12 \angstrom\ as most of them from the nearest spins in the Co kagome unit [Fig.3(b)]. Finally, we estimate the spin anisotropy energy by calculating the energy difference when rotating all local moments from ground state to the same angle within the $xoz$ plane or $xoy$ plane, which is about 0.6 meV at most between two geometries and consistent with the magnetization results [inset of Fig.3(b)] \cite{ekliu2018,jlshen2019}.

\subsection{Spin wave gaps}

The 3D spin interactions persisting both below and above $T_C$ with large stiffness suggest Co$_3$Sn$_2$S$_2$ has a moderate correlation and strong itinerancy to support the dispersive excitations against temperature. Hence the Weyl fermions related to the conduct electrons are very likely to contribute to the low-energy spin excitations. As the measurements on the temperature dependent dispersions of spin waves are extremely time consuming in triple-axis neutron experiments, we turns to map out the spin wave gap below $T_C$. In Fig.4(a), with a energy resolution $\Delta E =1 \sim 2$ meV, a clear reduction of spin excitations intensity below 4 meV can be found at $T=8$ K when we compare with $T=200$ K and 300 K data. With much higher resolution ($\Delta E \approx 0.1$ meV) measurements, the spin wave gap is precisely determined as a full gap below 2.3 meV at $T=4$ K [Fig.4(b)], where there is no excitation signal at $E=2$ meV but a tiny peak at $E=2.5$ meV [Fig.4(c)]. It gradually closes when warming up and finally disappears above $T_C$ [Fig.4(b)-(e)], giving a well-defined gaussian peak at $E=0.3$ meV and $T=175$ K [Fig.4(d)]. The spin wave gap is much larger than the estimated spin anisotropy energy, possible addition contributions can be from Weyl fermions.

Theoretically, the spin wave gap in magnetic WSMs at $q=0$ is given by $E_g=\mathcal{K}/\alpha$, where the coefficient $\alpha$ is the integrand represent of the Berry phase and $\mathcal{K}$ measures the spin anisotropy energy from SOC \cite{sitoh2016}(See Supplemental Materials). $\alpha$ has two contributions: one is $\alpha_0 =c_0/\langle S^z\rangle$ from the finite contribution without SOC, and the other is $\alpha_1=\lambda\sigma_{xy}$, which comes from the existence of Weyl nodes and directly relates with $\sigma_{xy}$ \cite{sitoh2016}. The parameter $\lambda$ is determined by the shape of Weyl cones. By introducing the normalization factors $M_0$ (saturated moment) and $\sigma_0=e^2/ha_0$ ($a_0$ is the lattice constant), the temperature dependence of $E_g$ can be described by the following phenomenological equation:
\begin{equation}
E_g(T)=\frac{aM(T)/M_0}{1+b(M(T)/M_0)(\sigma_{xy}(T)/\sigma_0)}.
\end{equation}
Here $a=\mathcal{K}\langle S^z\rangle_0/c_0$ and $b=\lambda\sigma_0\langle S^z\rangle_0/c_0$ are nearly temperature independent constants. Therefore, with the contribution from Weyl fermions ($b\neq0$), $E_g(T)$ probably deviates from the behavior of magnetic order parameter $M(T)$ for the involvement of $\sigma_{xy}(T)$. Although the temperature dependence of the gap shows an order-parameter-like behavior, accompanied by an abrupt decrease above $T_C$ for the integrated dynamic susceptibility $\chi''$ from $E_g$ to 4 meV [inset of Fig.4(f)], the direct fitting based on $E_g(T)=E_0(1-T/T_C)^{\beta}$ would give an unreasonable critical exponent $\beta=0.53$ with a large discrepancy for $\beta=0.21$ deduced from the actual order parameter in Fig. 1(g). Instead, we cannot fully describe $E_g(T)$ either by using fixed $\beta=0.21$ but ignore the contribution from Weyl fermions, as shown by the green dash line in Fig.4(f) (namely, $a=2.30$ meV and $b=0$ in Eq.3). Only when we consider the involvement of $\sigma_{xy}$ (which is $-\sigma^A_H$ in Fig.1(g)), we can well reproduce the $E_g(T)$, as shown by open circles in Fig.4(f), thus we have $a=0.93$ meV and $b=0.39$ in Eq.3. For comparison, the parameters $a=0.93$ meV and $b=0$ are also failed to fit the results. Therefore, the Weyl topology indeed strongly interplays with the spin dynamics. It should be noted that the coefficient $b$ here in Co$_3$Sn$_2$S$_2$ is positive and thus opposite to the case in SrRuO$_3$ with $b=$-9.5 or -4.98 (assuming $\sigma_{xy}$ is negative) \cite{sitoh2016,kjenni2019}, which depends on the detailed shapes of Weyl cones in these two compounds.

\section{Conclusions}

In summary, we have performed neutron scattering experiments on the recently verified magnetic WSM Co$_3$Sn$_2$S$_2$. In contrast to its quasi-2D lattice structure, the spin excitations both in the ferromagnetic and paramagnetic states are revealed with 3D characters. Theoretical calculations suggest comparable in-plane and out-of-plane magnetic exchange interactions, further estimation is consistent with the experimental $T_C$ and large spin stiffness. We have precisely determined a full spin wave gap and its temperature dependence below $T_C$, which is much larger than the spin anisotropy energy and certainly affected by the AHC triggered by the Weyl fermions. Our results give basic knowledge about the spin dynamics in Co$_3$Sn$_2$S$_2$ and solid evidences about how it interplays with the topological electron states.

{\bf Acknowledgments}

This work is supported by the National Key Research and Development Program of China (Grants Nos.: 2017YFA0303103, 2017YFA0302900, 2016YFA0300502, 2017YFA0206303), the National Natural Science Foundation of China (Grants Nos.: 11974392, 11974394, 11822411, 51722106, 11674372, 11774399, 11961160699 and 12061130200), the Strategic Priority Research Program (B) of the Chinese Academy of Sciences (CAS) (Grants Nos.: XDB07020300, XDB25000000), and Beijing Natural Science Foundation (Grants Nos.: JQ19002, Z180008, Z190009). E. L. and H. L. is grateful for the support from the Youth Innovation Promotion Association of CAS (Grants Nos.: 2013002, 2016004).

\clearpage
\begin{center}
{\bf SUPPLEMENTARY MATERIALS}

{\bf Spin Excitations and Spin Wave Gap in the Ferromagnetic Weyl Semimetal Co$_3$Sn$_2$S$_2$}

{\bf I. SAMPLE CHARACTERIZATION}
\end{center}

 The single crystals of Co$_3$Sn$_2$S$_2$ can be grown by Bridgeman, self-flux and chemical vapor transportation methods \cite{xlin2012s,wschnelle2013s,makassem2015s,qwang2018s,ekliu2018s}. We have successfully grown Co$_3$Sn$_2$S$_2$ sing crystals by flux method using Sn-Pb mixture as flux. The starting materials of Co, Sn, S and Pb in molar ratio of Co : S : Sn : Pb = 12 : 8 : 35 : 45 were put in Al$_2$O$_3$ crucibles sealed in quartz tubes. The quartz tubes were slowly heated to 400 $^{\circ}$C over 6 h and held on there over 6 h. Then further heating to 1100 $^{\circ}$C over 6 h was performed with a stay for 6 h. Then the melt was cooled down slowly to 700 $^{\circ}$C over 70 h. The flux was removed by rapid decanting and subsequent spinning in a centrifuge. The single crystal with typical sizes 2 - 5 mm and hexagonal edges were obtained. The naturally cleaving surface was $ab-$plane. We co-aligned the crystals by X-ray Laue reflection method with incident beam along $c$-axis on thin aluminium plates using \emph{CYTOP} hydrogen-free glue. The neutron scattering plane was defined by $[H, 0, 0] \times [0, 0, L]$  (Fig. S1(a)). With a clear hexagonal pattern and bright spots along $[H, 0, 0]$ direction (Fig. S1(b)), we were able to align about 200 pieces of crystals with total mosaic about 4$^{\circ}$ and total mass about 1.5 grams for Taipan experiment. The crystalline quality was also examined by single crystal X-ray diffraction (XRD) on a \emph{SmartLab} 9 kW high resolution diffraction system with Cu K$_{\alpha}$ radiation($\lambda$ = 1.5406 \angstrom) at room temperature ranged from $10\degree$ to $100\degree$ in reflection mode. Due to the symmetry restriction of $R\overline{3}m$ space group, only (0, 0, $3l$) peaks of X-ray diffraction could be found, where the sharp peaks indicated high quality of $c$-axis orientation of our samples (Fig. S1(c)). For the measurements of spin wave gaps at Sika, we have prepared a new large single crystal with mass about 7.6 grams, which was grown by self-flux method \cite{ekliu2018s}. The crystal was also aligned to $[H, 0, 0] \times [0, 0, L]$ scattering plane by  \emph{CYTOP} hydrogen-free glue and aluminium wires on a plate.

\begin{center}
{\bf II. NEUTRON SCATTERING EXPERIMENTS}
\end{center}

Neutron scattering experiments were carried out using a thermal triple-axis spectrometer Taipan and a cold triple-axis spectrometer Sika at ACNS, ANSTO, Australia, with fixed final energy $E_f=$ 14.87 meV and $E_f=$ 5 meV, respectively \cite{sdanilkin2009s,cmwu2016s}. First, the ferromagnetic order parameter was measured by elastic neutron scattering on a Bragg peak at $Q=$ (1, 0, 1) with weak nuclear scattering. The integrated intensity of (1, 0, 1) peak shown in Fig.1(f) indicates a clear ferromagnetic transition at $T_C=$ 175 K. The magnetically ordered moment of the ferromagnetic order can be estimated by comparing the magnetic scattering below $T_C$ and the nuclear scattering above $T_C$. In elastic neutron scattering, the nuclear peak intensity $I_N$ and magnetic peak intensity $I_M$ are determined by \cite{gshiranes,dlgong2018s,rzhang2015s}:
\begin{equation}
I_N=AN_N(2\pi)^2/V_N \times(\mid F_N(Q)\mid)^2/\sin(2\theta_N),
\end{equation}
and
\begin{equation}
I_M=AN_M(2\pi)^2/V_M \times(\mid F_M(Q)\mid)^2/\sin(2\theta_M),
\end{equation}
respectively. For simplicity, here we ignore the orbital magnetism and spin-orbit coupling. For a ferromagnet with ordered magnetic moments, the magnetic unit cell is identical to the nuclear unit cell, then we have $N_M=N_N$, $V_M=V_N$ and $\theta_M=\theta_N$. The intensity difference only comes from the nuclear structure factor $F_N(Q)$ and the magnetic structure factor $F_M(Q)$, the latter can be expressed as $F_M(Q)= (\gamma r_0 / 2)\times f_M(Q)\times m\times\sin{\eta}\times\sum e^{i{\bf Q}\cdot {\bf d}}$, where $\gamma$ is the gyromagnetic ratio of neutron, $r_0$ is the classical electron radius ($\gamma r_0/2=0.2695 \times 10^{-12}$ cm), $f_M(Q)$ is the magnetic form factor of Co, $m$ is the static moment, and $\eta$ is the angle between the spin and wave vector $\bf Q$.  For the measured $Q=(1, 0, 1)$, we have $F_N=3.97$ and $f_M=0.89$, $\sin{\eta}=0.943$, $\sum e^{i{\bf Q}\cdot {\bf d}}$=3. then we can estimate the ordered moment in the unit of $\mu_B$/Co via:
\begin{equation}
m=0.131\sqrt{I_M/I_N}\mid F_N\mid/\mid f_M\mid=0.6\sqrt{I_M/I_N}.
\end{equation}
 Considering the linear background above $T_C$ in Fig. 1(e), the ordered moment at base temperature $T=8$ K is estimated about $0.28\pm 0.02 \mu_B$ per Co atom. The leakage of high order neutrons from pyrolytic graphite filter and the neutron absorption from Co atoms may give uncertainties of such estimation as well.  Thus it agrees with previous results from transport measurements (about 0.3$\mu_B$/Co) within experimental errors \cite{makassem2015s,qwang2018s,ekliu2018s,pvaqueiro2009s,makassem2016as,makassem2016bs}. From the temperature dependence of peak intensity at $Q=$(1, 0, 1) in Fig.1(f), we can't find any anomaly around $T_A=136$ K, which may be related to another in-plane antiferromagnetic order in this compound \cite{makassem2017s,zguguchia2019s}.

For inelastic neutron scattering experiment at Taipan, constant-energy scans ($Q-$scans) were performed from $E=2$ meV to 18 meV around $Q=(0, 0, 3)$ both along $H$ and $L$ directions. For all $H-$scans ranged [-0.52, +0.52], the magnetic excitation signals are gaussian-like (either single peak or two peaks) with a flat background, the excitation intensity $S(Q, \omega)$ can be obtained by simply subtracting the flat background determined by the data outside the peak from $-0.52\leqslant H \leqslant -0.3$ and $+0.3\leqslant H \leqslant +0.52$ (Fig. S2(a)(c)). For $L-$scans ranged [1.5, 4.5], we have found the background are non-linear due to phonon scattering from both sample and the aluminium holder, thus we have measured the background by identical scans along $[\pm0.3, 0, L]$ and further testified it by additional scans along $[0.5, 0, L]$ (Fig. S2(b)(d)), where there is no magnetic scattering signal from $H-$scan results. The final excitation intensity $S(Q, \omega)$ of $L-$scans are obtained by point-to-point subtraction between the $[0, 0, L]$ scans and the average of $[\pm0.3, 0, L]$ background scans. For comparison, we mark the instrument resolution calculated by ResLib in Fig. S2, where the peak width of $L$ scans are much broader than the resolution, probably due to its short range dynamic spin correlations. We present all raw data of $Q-$scans in Fig. S3 and Fig. S4 after the subtracting the background. For each $H$ scan, we have fitted the data by a gaussian function with two symmetric peaks as following:
\begin{equation}
I= I_0[e^{-(H-\delta)^2/2\sigma^2}+e^{-(H+\delta)^2/2\sigma^2}].
\end{equation}
For each $L$ scan, the intensity of two gaussian peaks are slightly asymmetric due to the normalization effect from magnetic form factor $f(\mid Q\mid)$:
\begin{equation}
I= I_1[e^{-(L-3-\delta)^2/2\sigma^2}]+I_2[e^{-(L-3+\delta)^2/2\sigma^2}],
\end{equation}
where the ratio between $I_1$ and $I_2$ are solely determined by the form factor $f(\mid Q\mid)$ \cite{dprice2015s}. The symmetric peak positions are determined by $\delta$, which mark the dispersions in Fig. 2(e) and (f). Although there are some supurious signals for $E=2$ meV data and a upturn at small $L$ side from the main beam contamination, no well-defined peaks can be found both for $H-$scan and $L-$scan at $T=8$ K, which means the spin wave gap is a full gap below 2 meV. The spin excitations in the form of a small peak emerge at $E=2.5$ meV which becomes gaussian-like at $E=3$ meV.

We have also found strong phonon scattering signals for $H-$scans above $E=$14 meV outside the inelastic magnetic peak, which heavily contaminate the 15 meV data and make it is impossible to fit. For this reason, we won't discuss the relative intensity change upon energy above 14 meV. We have carried out additional scans from $H=0.52$ to $H=1$, and $E=15$ meV to 20 meV. Indeed, the phonon scattering grows up quickly in the high $Q$ region, and its dispersion (probably belong to two optical branches) seems to follow $H^2$ relationship (Fig. S5), but much broader than the magnon dispersion.

To precisely determine the spin wave gap below $T_C$, we have further performed high resolution inelastic neutron scattering experiments at Sika on a large single crystal. With the energy transfer covering from 0.3 meV to 4 meV, we have measured the spin excitations from $T=4$ K to $T=300$ K. As shown in Fig.S6, here the magnetic scattering signals are measured at $Q=(0, 0, 3)$ and the backgrounds are measured by identical countings at $(\pm0.5, 0, 3)$. By point-to-point subtraction, we can obtain the relative intensity for each energy scans. A clear cutoff for all spin excitations below $T_C$ can be found, suggesting a spin wave gap develops in the ferromagnetic state, where the spin excitation intensity is nearly zero below $E_g$. Above $T_C$, the gap is fully closed, and only a upturn at low energy remains. The local susceptibility $\chi''(Q, \omega)$ shown in Fig.4(e) is obtained from the data in Fig. S6(b1)- (b14) after removing the Bose population factor $1/(\exp(h\omega/k_BT)-1)$, while the total dynamic susceptibility $\chi''$ in the insert of Fig.4(f) is integrated from $E_g$ to 4 meV.

\begin{center}
{\bf III. CALCULATION METHOD OF SPIN DYNAMICS PARAMETERS}
\end{center}
Although the Co$_3$Sn$_2$S$_2$ compound is believed as an itinerant ferromagnet below 175 K, we can still estimate the effective exchange interactions by viewing the effective local moments on Co atoms. The semi-classical analogous to the Heisenberg Hamiltonian is written as:
\begin{equation}
  H = \sum_{<i,j>} J_{ij} {\bf e}_i {\bf e}_j.
\end{equation}
The unit vector ${\bf e}_i$ points to the same direction as the $i$th site magnetization. $J_{ij}$ stands for the exchange interaction energy between the $i$th and $j$th sites.
For such a classical system, one can get the total energy variation as a function of spin rotation angle $\pm\theta/2$ on site $i,j$ in the scope of the multiple-scattering Green function method, the leading order of $\theta$ is:
\begin{equation}
  \delta E_{ij} = \frac{1}{8\pi} \int^{E_f}dE\ {\rm Im}\ {\rm Tr}_{L} \{\hat{\Delta}_i \hat{T}_{ij,\uparrow}\hat{\Delta}_j \hat{T}_{ji,\downarrow} \} \theta^2.
\end{equation}
The expression of scattering path operator $T_{ij}$ is the fundamental equation of multiple scattering theory:
\begin{equation}
  \hat{T}_{ij} = (\hat{t}_{i}\delta_{ij} - \hat{G}_{ij})^{-1},
\end{equation}
and $\hat{\Delta}_{i} = \hat{t}^{-1}_{i,\uparrow} - \hat{t}^{-1}_{i,\downarrow}$ is the difference between the single site scattering matrix $\hat{t}_{i}$. The expression have already exclude the interaction of one site with the system, only consider the energy difference caused by $i,j$ site.

In this perturbation method, the total energy calculation can also be obtained by the use of Andersen's local force theorem, which tells us that the total energy variation coincides with the sum of one particle energy changes.
Thus, by rotating a little angle $\theta$ on the site 0 from the ground state, the energy difference is:
\begin{equation}
  \delta E_0 = \sum_{j} J_{0j}(1-\cos(\theta))\approx \frac{1}{2}J_0 \theta^2,
\end{equation}
here $J_0$ are defined as the effective exchange parameters for one site feels the interaction of the whole system:
\begin{equation}
  J_0 = \sum_{j} J_{0j}.
\end{equation}
When we rotate two sites by $\theta/2$, the energy difference is:
\begin{equation}
  \delta E_{ij} = J_{ij}(1-\cos(\theta))\approx \frac{1}{2}J_{ij}\theta^2.
\end{equation}
By comparing these two equations we have:
\begin{equation}
  J_{ij} = \frac{1}{4\pi}\int^{E_f}dE\ {\rm Im}\ {\rm Tr}_{L} \{\hat{\Delta}_i \hat{T}_{ij,\uparrow}\hat{\Delta}_j\hat{T}_{ji,\downarrow}\},
\end{equation}
for which the calculation can be implemented in the Wannier basis scheme.  Note that Wannier function is a natural choice of local basis sets, which coincides to the requirement of the exchange Hamiltonian \cite{dmmkorotin2015s}.

We calculated the interactions between one and nearest 400 Co atoms, the nearest and next nearest in-plane and out-plane interactions are listed below: $J_1= -2.37$ meV, $J_2= -0.16$ meV, $J_3= -0.02$ meV, $J_{c1}= -0.30$ meV, $J_{c2}= -0.75$ meV and $J_{c3}= -0.11$ meV [Fig.3(a)]. The strongest out-of-plane interaction is about one third of the nearest in-plane interaction, which is consistent with the experimental data for 3D spin waves. To see how the exchange interactions change with distance, we summed over all the contributions from atoms have same distance to the central Co atom, the results are shown in Fig. 3(b). The biggest contribution comes from the triangular structure in Co kagome layer as expected. All out-of-plane and in-plane interactions disappear once the distance reaches 12 \angstrom\ due to the local moment nature of our model.

The calculation results can be verified by estimating the Curie temperature $T_C$. Under the mean field approximation, the Curie temperature of a ferromagnetic systems is given by:
\begin{equation}
  T_C = \frac{2J_0\langle M^2 \rangle}{3k_B \langle M\rangle^2}.
\end{equation}
For classical ferromagnetic systems we have $\langle M^2\rangle /\langle M\rangle^2 = 1$ \cite{liechtenstein1987s}, thus we can estimate the Curie temperature by $T_C = 2J_0/3k_B$, which is 167 K and fairly close to the experimental value. We can further estimate the spin wave stiffness, as it is determined by the following expression:
\begin{equation}
  D_{\alpha\beta} = \frac{2}{M} \sum_j J_{0j}R_{j\alpha}R_{j\beta}
\end{equation}
where $R_{j\alpha}$ means the $\alpha$-direction component of lattice vector ${\bf R}_{j}$. By using the calculated $J_{ij}$, we get $D_{xx} = 945$ meV\angstrom$^2$, $D_{yy}=833$ meV\angstrom$^2$ and $D_{zz}=656$ meV\angstrom$^2$. The in-plane stiffness is closed to the experiment data, but out-of-plane stiffness indicates that the calculation may overestimate the exchange coupling along $z$ axis.

In principle, we can estimate the spin anisotropy energy by using a general model:
\begin{equation}
  H = J\sum_{i,j}(S_i^x S_j^x + S_i^y S_j^y + A S_i^z S_j^z).
\end{equation}
When $A = 1$, the isotropy model gives degenerate ground states for all directions, and there is no gap in the spin wave spectrum. The nontrivial $A$ lifts the degeneracy and the excitation spectrum acquire a gap $|J|zSA$. In real materials the anisotropy terms are much more complicated, but we can check its existence instead by rotating all local moments from ground state to the same angle, and compare the energy difference. By rotating the local moment on the $xoz$ plane, the system acquire energy gradually, and reach the peak when the local moment point to $x$ axis. The energy difference to the ground state is about 0.6 meV. As a comparison, the energy difference is two orders of magnitude less when we rotating the local moment in the $xoy$ plane [insert of Fig.3(b)]. These results indicating an non-negligible anisotropy term on $S_i^z S_j^z$, which may contribute to the spin wave gap.

\begin{center}
{\bf IV. SPIN WAVE GAP AND ITS TEMPERATURE DEPENDENCE IN MAGNETIC WEYL SEMIMETALS}
\end{center}
The spin wave excitation in a ferromagnetic metal is developed by Onoda {\it et al.} \cite{monoda2008s} and applied to SrRuO$_3$ case by Itoh {\it et al.} \cite{Itoh2016s}, lately confirmed by a single crystal neutron scattering experiment on SrRuO$_3$ \cite{kjenni2019s}. Here we repeat derivation of Eq.2, which is a general model and can be applied to Co$_3$Sn$_2$S$_2$ as well.

In the weak correlation limit, the spin wave dispersion is determined by the RPA-type spin susceptibility $\chi^{+-}({\bf q},\omega)$ of electrons as shown below \cite{tizuyama1963s}:
\begin{equation}
    \chi^{+-}({\bf q},\omega) = \frac{\chi_0^{+-}({\bf q},\omega)}{1 - U\chi_0^{+-}({\bf q},\omega)},
\end{equation}
where the bare susceptibility without electron-electron interaction is given as
\begin{equation}
    \chi_0^{+-}({\bf q},\omega) = \sum_k \frac{f(\varepsilon_\uparrow ({\bf k})) - f(\varepsilon_\downarrow ({\bf k}))}{\varepsilon_\uparrow ({\bf k})-\varepsilon_\downarrow ({\bf k+q})-\hbar\omega}.
\end{equation}
By solving the equation $1=U\chi_0^{+-}({\bf q},\omega)$, one can obtain the spin wave dispersion $\omega(q)$, and the electron correlation is determined by the energy distribution of up and down spins: $U = (\varepsilon_\downarrow ({\bf k}) - \varepsilon_\uparrow ({\bf k}))/(2\langle S_z \rangle)$. When the spin-orbit interaction in the electronic band structure is included, we then have a more complex form for $\chi_0^{+-}({\bf q},\omega)$:

\begin{equation}
\begin{aligned}
    \chi_0^{+-}({\bf q},\omega) = &\sum_{k,n,m} \frac{f(\varepsilon_n ({\bf k})) - f(\varepsilon_m ({\bf k+q}))}{\varepsilon_n ({\bf k})-\varepsilon_m ({\bf k+q})-\hbar\omega}\\&\times\langle n{\bf k}|S^+|m{\bf k+q}\rangle\langle m{\bf k+q}|S^-|n{\bf k}\rangle,
\end{aligned}
\end{equation}

where $n,m$ are the band indices including the pseudospin. the zero-th term from the expansion with respect to $\omega$ and $q$ of above equation gives the spin anisotropy $\mathcal{K}=\chi^{+-}(0,0)-\chi^{zz}(0,0)$, which is basically independent of $T$ near $T_C$ and adds the gap $E_g=\mathcal{K}\langle S^z\rangle$ to the spin wave dispersion $\omega(q)$. After applying the Vienna ab-initio simulation package (VASP) \cite{gkresse1996s} to the first principle calculation, we have found such anisotropy can solely induce a spin wave gap about 1.45 meV, which is higher than the former estimated spin anisotropy energy (0.6 meV) in the insert of Fig.3(b) but still lower than the experimental value (2.3 meV).

Now we consider the spin wave gap in a magnetic Weyl semimetal with contributions from Weyl fermions. The effective action $A$ for the spin fluctuation field $S^{\alpha}$ is obtained by integrating over the electronic degrees of freedom as \cite{monoda2008s}:
\begin{equation}
A=\sum_{q,\omega,\alpha\beta=x,y,z} \chi^{\alpha\beta}(q,\omega)S^{\alpha}(q,\omega)S^{\beta}(-q,-\omega),
\end{equation}
where $\chi^{\alpha\beta}(q,\omega)=\langle \sigma^{\alpha}(q,\omega)\sigma^{\beta}(-q,-\omega)\rangle$ is the correlation function of the conduction electron spin $\sigma^{\alpha}$. The spin wave gap locally emerges at $q=0$, then we have:
\begin{equation}
\begin{aligned}
A_0=&\frac{1}{2}\int dt\{\alpha[(dS^x/dt)S^y-(dS^y/dt)S^x]\\&-\mathcal{K}[(S^x)^2+(S^y)^2]\},
\end{aligned}
\end{equation}
where the coefficient $\alpha$ is the integrand represent of the Berry phase: $\alpha=\chi^{xy}(0,\omega)/i\omega=\langle \sigma^{x}, \sigma^{y}\rangle/i\omega$.  $\alpha$ consists of two terms: the first term $\alpha_0=c_0/\langle S^z\rangle$ contributed from the bands not largely influenced by the spin-orbit interaction, and the second term $\alpha_1=\lambda\sigma_{xy}$ contributed from the Weyl fermions and proportional to the anomalous Hall conductivity ($\sigma_{xy}$) \cite{Itoh2016s}. $\lambda$ can be deduced from the model Weyl Hamiltonian written as:
\begin{equation}
    H = \sum_{ab} g_{ba}(k_b-k_{0b})\sigma^a.
\end{equation}
Here, ${\bf k}_0$ represents the location of the Weyl points in momentum space. The current operator $j_b$ is related to the spin operator $\sigma^a$ by $j_b = e\sum_{a} g_{ba}\sigma^a$ after replacing ${\bf k}$ with ${\bf k}+e{\bf A}$ in above Hamiltonian. In detail, we have
\begin{equation}
\begin{aligned}
    \langle \sigma^{x}, \sigma^{y}\rangle &=\frac{1}{e^2}(g^{-1}_{xx}g^{-1}_{yy}-g^{-1}_{yx}g^{-1}_{xy})\langle j_x, j_y\rangle\\
                                          &+\frac{1}{e^2}(g^{-1}_{yx}g^{-1}_{zy}-g^{-1}_{zx}g^{-1}_{yy})\langle j_y, j_z\rangle\\
                                          &+\frac{1}{e^2}(g^{-1}_{zx}g^{-1}_{xy}-g^{-1}_{xx}g^{-1}_{zy})\langle j_z, j_x\rangle,
\end{aligned}
\end{equation}
and there is no contributions coming from $\sigma_{xz}$ and $\sigma_{yz}$ in Co$_3$Sn$_2$S$_2$ case, which yields $\langle j_y, j_z\rangle=\langle j_z, j_x\rangle=0$. Then we have $\lambda = e^{-2}(g_{xx}^{-1}g_{yy}^{-1}-g_{yx}^{-1}g_{xy}^{-1})$, which is slightly different from the SrRuO$_3$ case \cite{Itoh2016s}. By taking the variational
derivative of $A_0$ with respect to $S^x$ and $S^y$, one can derive the equation of motion and obtain the spin wave gap as:
\begin{equation}
 E_g=\mathcal{K}/(c_0/\langle S^z\rangle+\lambda\sigma_{xy}).
\end{equation}
By introducing the normalization factors $M_0$ (saturated moment at $T=0$ K) and $\sigma_0=e^2/ha_0$ ($a_0$ is the lattice constant), the temperature dependence of $E_g$ can be alternatively described by the following equation:
\begin{equation}
E_g(T)=\frac{aM(T)/M_0}{1+b(M(T)/M_0)(\sigma_{xy}(T)/\sigma_0)},
\end{equation}
with $a=\mathcal{K}\langle S^z\rangle_0/c_0$ and $b=\lambda\sigma_0\langle S^z\rangle_0/c_0$ (note $\langle S^z(T)\rangle/\langle S^z\rangle_0=M(T)/M_0$). Therefore, with the contribution from Weyl fermions ($b\neq0$), $E_g(T)$ probably deviates from the magnetic order parameter $M(T)/M_0=(1-T/T_C)^{\beta}$.\\
\\
\\
\\
\\
\\

\newpage
\begin{figure*}[t]
\renewcommand\thefigure{S1}
\includegraphics[width=0.6\textwidth]{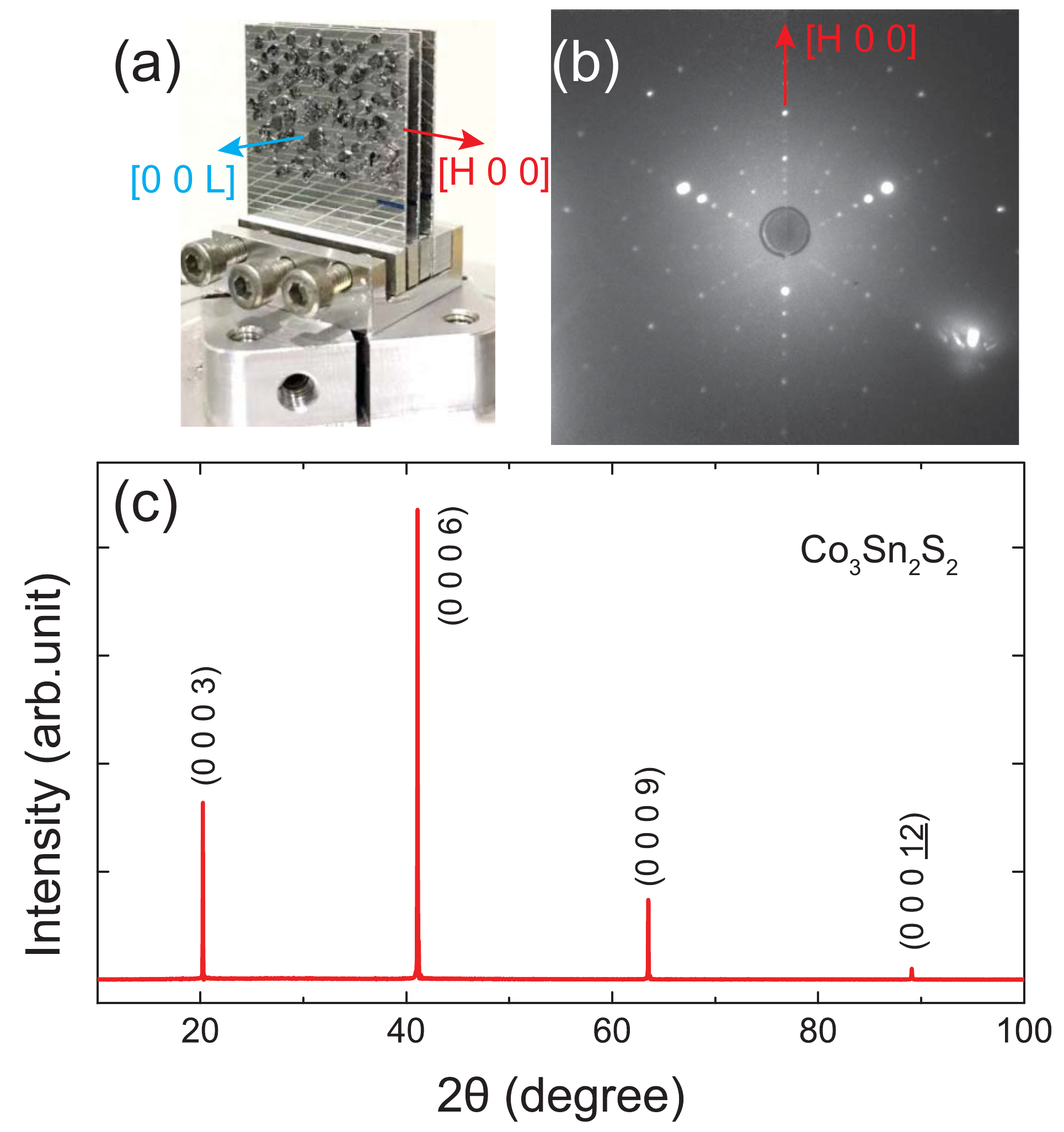}
\caption{
(a) Photos of the co-aligned Co$_3$Sn$_2$S$_2$ crystals in $[H, 0, L]$ scattering plane for our neutron experiments. (b) X-ray Laue reflection pattern of Co$_3$Sn$_2$S$_2$ single crystal at room temperature. The high symmetry direction $[H, 0, 0]$ is indicated by the red arrow. (c) X-ray diffraction pattern for Co$_3$Sn$_2$S$_2$ single crystal with incident beam along $c-$axis.
}
\end{figure*}

\begin{figure*}[t]
\renewcommand\thefigure{S2}
\includegraphics[width=0.7\textwidth]{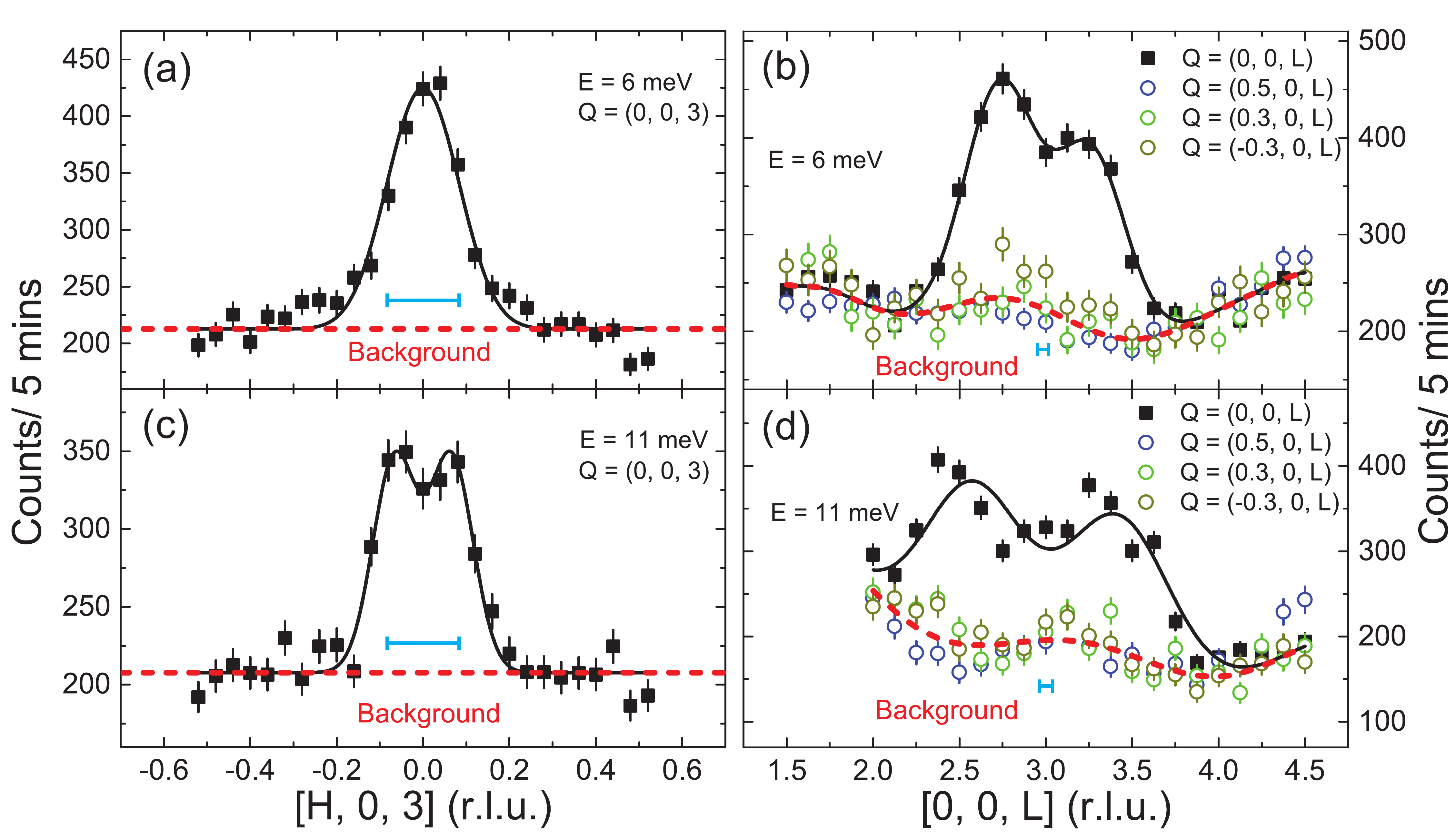}
\caption{Raw data for constant-energy scans at $E=6$ and 11 meV, $T=8$ K. (a) and (c) The background (red dashed lines) for $H-$scans are flat for $-0.52\leqslant H\leqslant +0.52$. (b) and (d) The background (red dashed lines) for $L-$scans are determined by identical scans along $[\pm0.3, 0, L]$ and testified by additional scan along $[0.5, 0, L]$, where there is no magnetic scattering signal from $H-$scan results. The horizontal bars are the instrument resolution calculated by ResLib.
 }
 \end{figure*}

\begin{figure*}[t]
\renewcommand\thefigure{S3}
\includegraphics[width=0.8\textwidth]{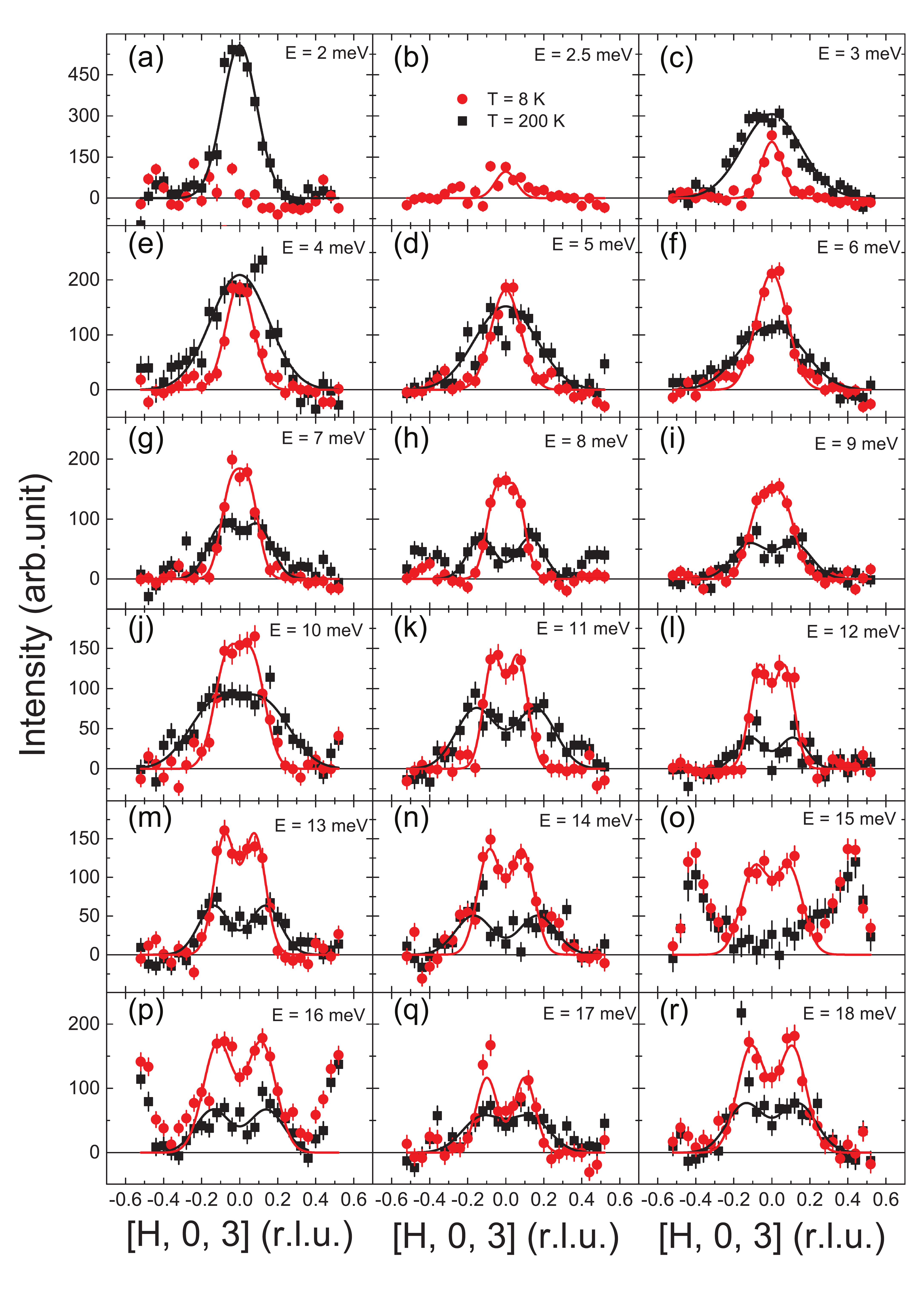}
\caption{Two-gaussian-fitting results for all $[H, 0, 3]$ scans both at $T=8$ K and 200 K. A flat background is already subtracted before the fitting, and the phonon scattering at high $Q$ is ignored above 14 meV.
 }
\end{figure*}

\begin{figure*}[t]
\renewcommand\thefigure{S4}
\includegraphics[width=0.8\textwidth]{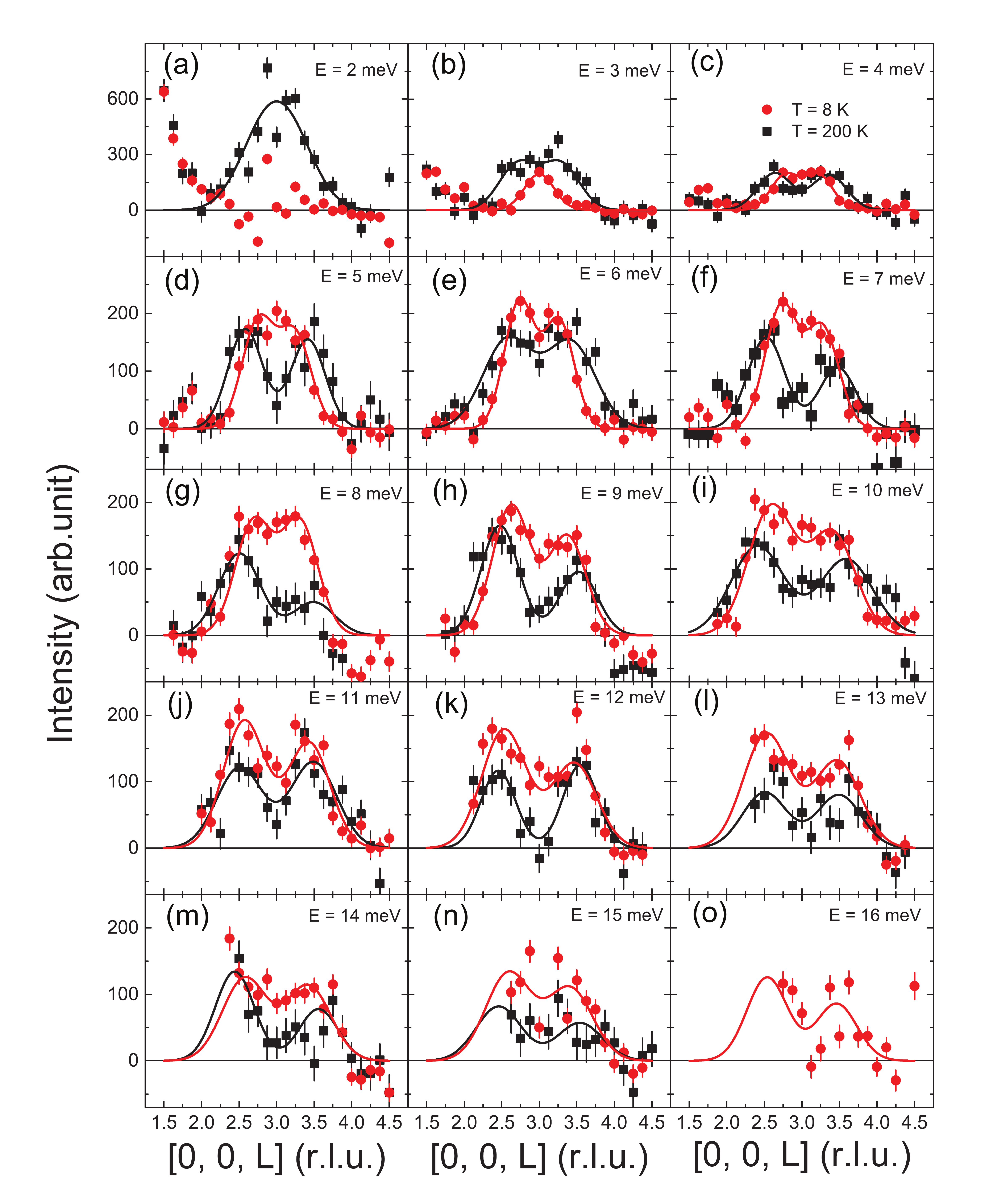}
\caption{
Two-gaussian-fitting results for all $[0, 0, L]$ scans both at $T=8$ K and 200 K. All background are subtracted by identical scans along $[\pm0.3, 0, L]$, and the upturn at low $Q$ due to the tail of main beam is ignored for $E=2$, 3, 4, 5 meV.
 }
\end{figure*}

\begin{figure*}[t]
\renewcommand\thefigure{S5}
\includegraphics[width=0.8\textwidth]{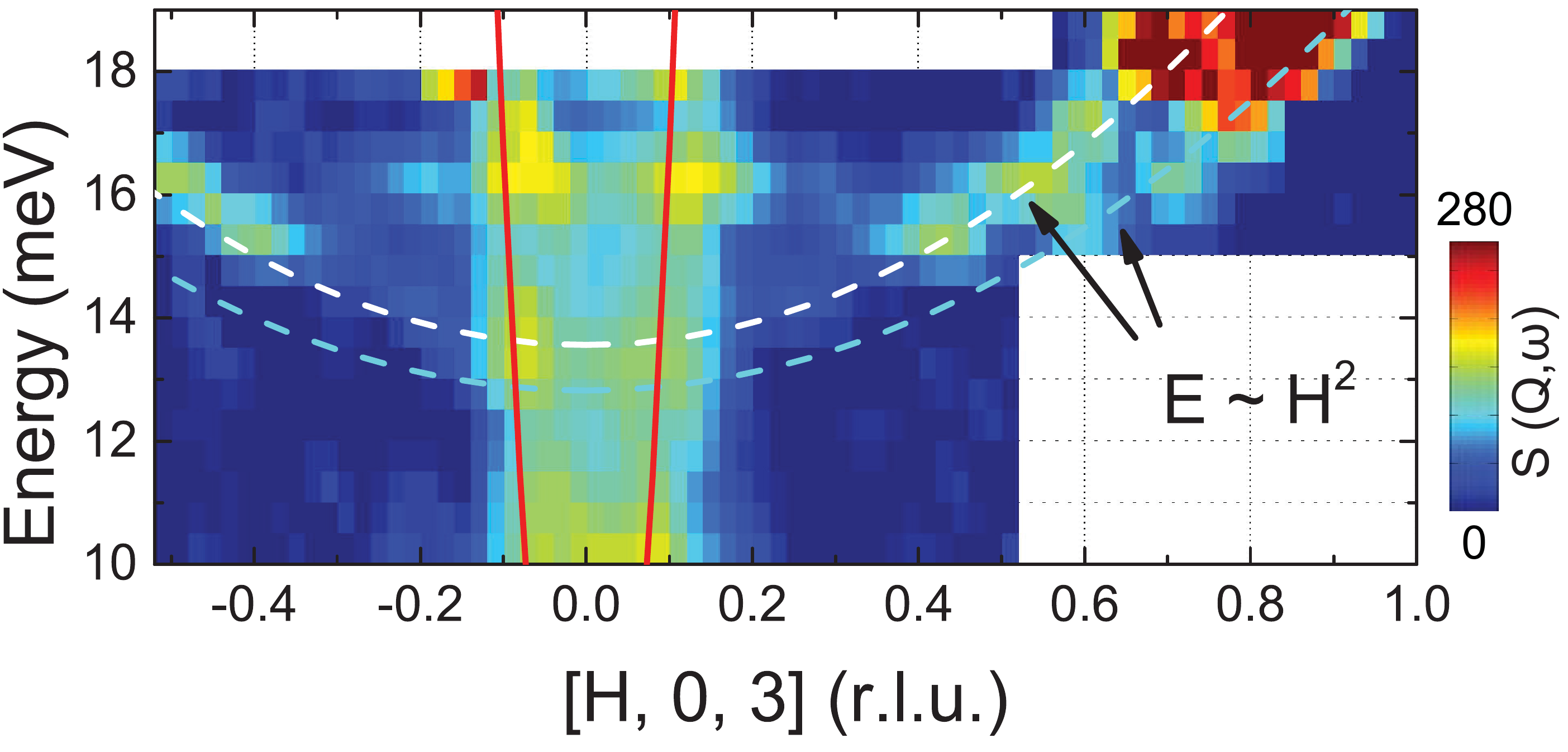}
\caption{Color mapping for the inelastic scattering from $E=10$ meV to 18 meV. The dashed lines mark the phonon dispersions with $H^2$ dependence.
 }
\end{figure*}

\begin{figure*}[t]
\renewcommand\thefigure{S6}
\includegraphics[width=0.95\textwidth]{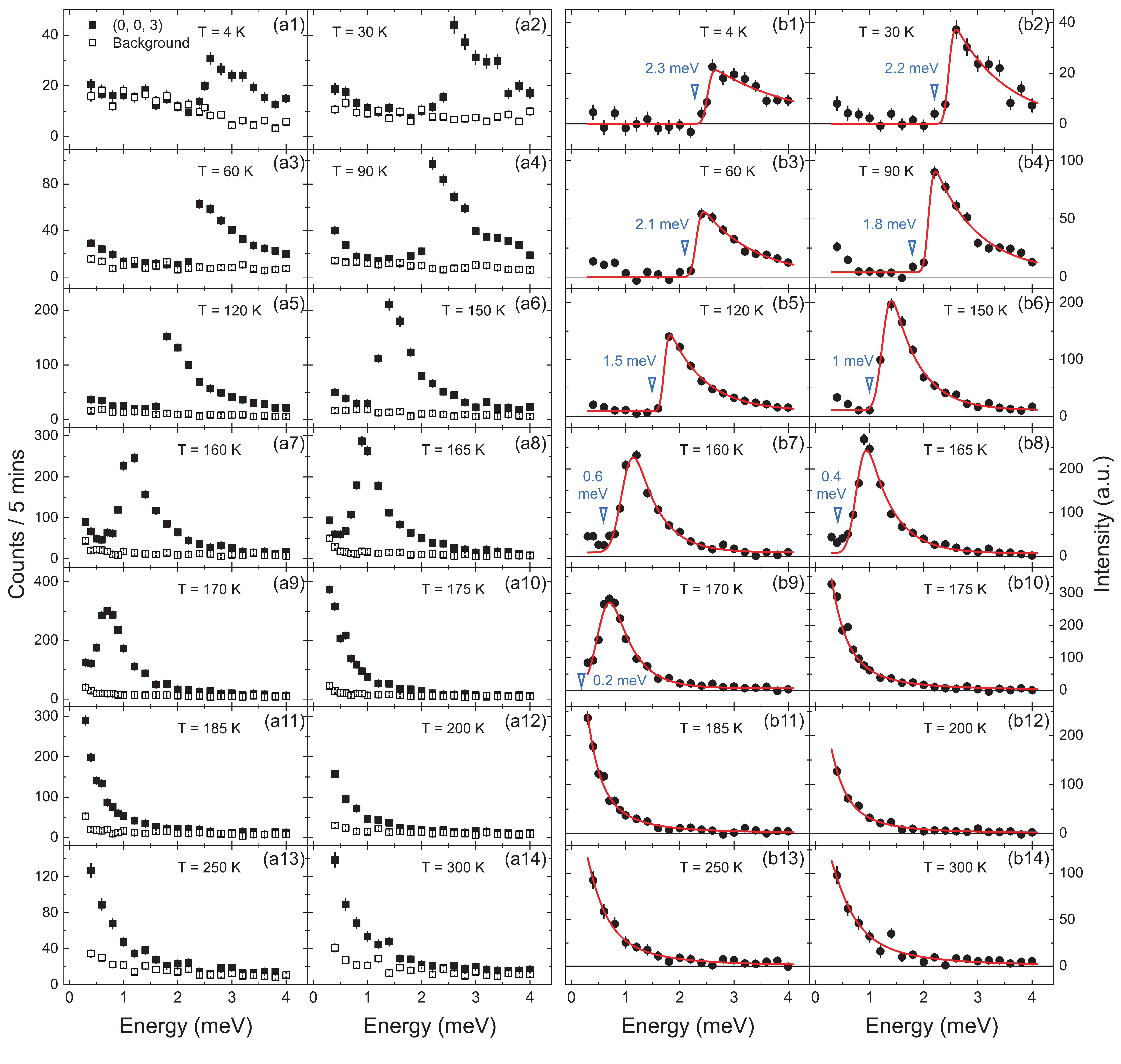}
\caption{ High resolution energy scans measured at the cold triple-axis spectrometer Sika.
(a1) - (a14) Raw data for energy scans from $T=4$ K to $T=300$ K, where the magnetic scattering signals are measured at $Q=(0, 0, 3)$ and the backgrounds are measured by identical countings at $(\pm0.5, 0, 3)$.
(b1) - (b14) Magnetic excitation intensities after subtracting the backgrounds from $T=4$ K to $T=300$ K, where the spin wave gap is marked in each panel below $T_C=175$ K.
 }
\end{figure*}

\end{document}